%% file: LHCP_2014_-_ewkZjj.tex
\pdfoutput=1
\documentclass[10pt]{article}
\usepackage{graphicx}
\usepackage{xspace}
\usepackage{amsmath}

\newcommand{\Madgraph}{M\scalebox{0.8}{AD}G\scalebox{0.8}{RAPH}\xspace}

\newcommand{\Mcfm}{\scalebox{0.8}{MCFM}\xspace} 
\newcommand{\pt}{\ensuremath{p_{\mathrm{T}}}}

\def\Title#1{\begin{center} {\Large #1 } \end{center}}
\def\Author#1{\begin{center}{ \sc #1} \end{center}}
\def\Address#1{\begin{center}{ \it #1} \end{center}}

\newcommand\pubblock{\rightline{\begin{tabular}{l} Proceedings of the Second Annual LHCP\\ \pubnumber\\
         \pubdate  \end{tabular}}}

\newenvironment{Abstract}{\begin{quotation} \begin{center} 
             \large ABSTRACT \end{center}\bigskip 
      \begin{center}\begin{large}}{\end{large}\end{center} \end{quotation}}

\newenvironment{Presented}{\begin{quotation} \begin{center} 
             PRESENTED AT\end{center}\bigskip 
      \begin{center}\begin{large}}{\end{large}\end{center} \end{quotation}}

\input econfmacros.tex

\usepackage{color}

\newcommand\pubnumber{ CMS-CR-2014/185 }

\newcommand\pubdate{\today}

\def\affiliation{
On behalf of the CMS Collaboration, \\
Department of Elementary Particle Physics \\
Universiteit Antwerpen, 2020 Antwerp, Belgium}

\begin{document}

\large
\begin{titlepage}
\pubblock

\vfill
\Title{Measurement of the electroweak production of a Z boson and two jets in proton-proton collisions at 7 and 8 TeV}
\vfill

\Author{Tom Cornelis}
\Address{\affiliation}
\vfill
\begin{Abstract}
A measurement of the electroweak production cross section of a Z boson and two jets is performed using proton-proton collision data at centre-of-mass energies of 7 and 8 TeV. 
The data is collected by the CMS experiment at the LHC with an integrated luminosity of 5 and 19 fb$^{-1}$, respectively. 
A study on the hadronic activity between the two tagging jets is also presented.

\end{Abstract}
\vfill

\begin{Presented}
The Second Annual Conference\\
 on Large Hadron Collider Physics \\
Columbia University, New York, U.S.A \\ 
June 2-7, 2014
\end{Presented}
\vfill
\end{titlepage}
\def\thefootnote{\fnsymbol{footnote}}
\setcounter{footnote}{0}
%

\normalsize 


\section{Introduction}
\begin{figure}[htb]
   \centering
   \def\feynmannHeight{2.5cm}
   \includegraphics[height=\feynmannHeight]{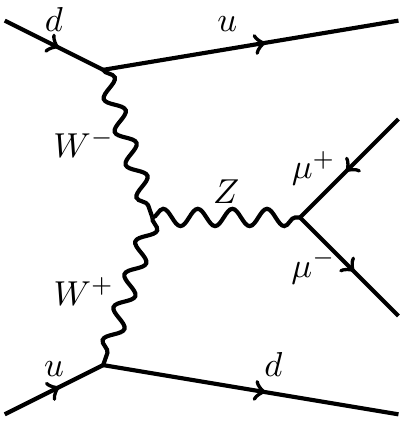}
   \includegraphics[height=\feynmannHeight]{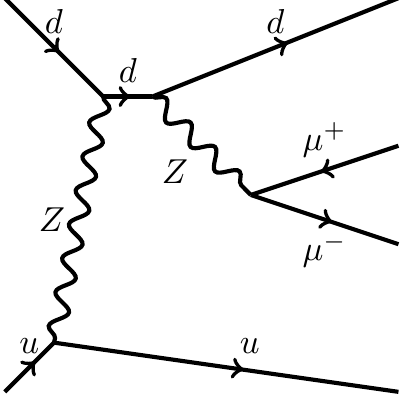}
   \includegraphics[height=\feynmannHeight]{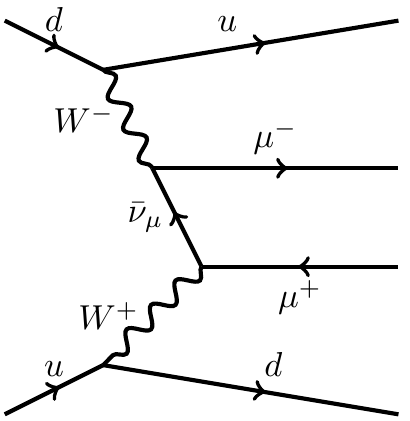}
   \caption{Representative diagrams for the electroweak $Z\rm jj$ production processes: VBF, bremsstrahlung and multi-peripheral.}
 \label{fig:feynmann}
 \end{figure}

 In proton-proton collisions at the LHC, the dominant source of events with the $lljj$ final state is through mixed electroweak and strong processes of order $O(\alpha_{EW}^2\alpha_{QCD}^2)$, 
 also known as Drell-Yan plus two jets. The pure electroweak production of a $Z$ boson in association with two jets is a rarer process.
 Different classes of pure electroweak $O(\alpha_{EW}^4)$, $lljj$ processes are possible: vector boson fusion (VBF), bremsstrahlung and multi-peripheral, as shown in Figure \ref{fig:feynmann}.

 Because of large gauge cancellations between them, these classes cannot be separated in data. The search for these pure electroweak processes exploits some distinctive event properties:
 \begin{itemize} 
     \item a central $Z$ decay associated with two energetic forward-backward light-quark jets,
     \item a large dijet $\eta$ separation and a large dijet invariant mass $M_{jj}$,
     \item colour exchange suppression between the tagging quark jets.
 \end{itemize} 
 The results of this analysis pave the road for the more general study of vector boson fusion processes and for measurements of electroweak gauge couplings and vector boson scattering.

 Events are selected by requiring a muon or electron pair with opposite charge, in which both leptons 
 satisfy standard CMS quality and isolation criteria, a transverse momentum of at least 20 GeV and $|\eta| < 2.4$.
 The reconstructed dilepton mass is required to be within the 15 GeV (muon channel) or 20 GeV (electron channel) of the nominal $Z$ boson mass. 
 The two leading jets within $|\eta| < 4.7$ are selected as the tagging jet candidates. In the 7 TeV analysis, events are selected with tagging jets exceeding $p_T^{j_1,j_2} > 65,40$ GeV, while
 this criterion was relaxed to $p_T^{j_1,j_2} > 50,30$ GeV in the 8 TeV analysis.

\section{Background modelling}
 The standard simulation for Drell-Yan plus jets, the main background to the electroweak $jjll$ process, is based on \Madgraph and lacks higher order virtual corrections.
 Predictions using \Mcfm are therefore used to derive LO to NLO correction factors based on $M_{jj}$ and on the rapidity of the Z boson in the dijet mass frame ($y^*$).
 For the 8 TeV analysis, a data-driven approach is added. The production of a photon plus two jets is expected to resemble the Drell-Yan processes and can therefore be used
 to model the properties of the tagging jets. The photon $\pt$ is reweighted to the $Z$ boson $\pt$ in order to mitigate the differences induced by the specific $\gamma$ or Z sample. 
 Because the photon sample is affected by multijet production and high trigger prescales at low $p_T$, the photon or $Z$ is required to have a $\pt$ larger than 50 GeV. 

\section{Cross section measurement}
\begin{figure}
  \centering
  \includegraphics[width=0.3\columnwidth]{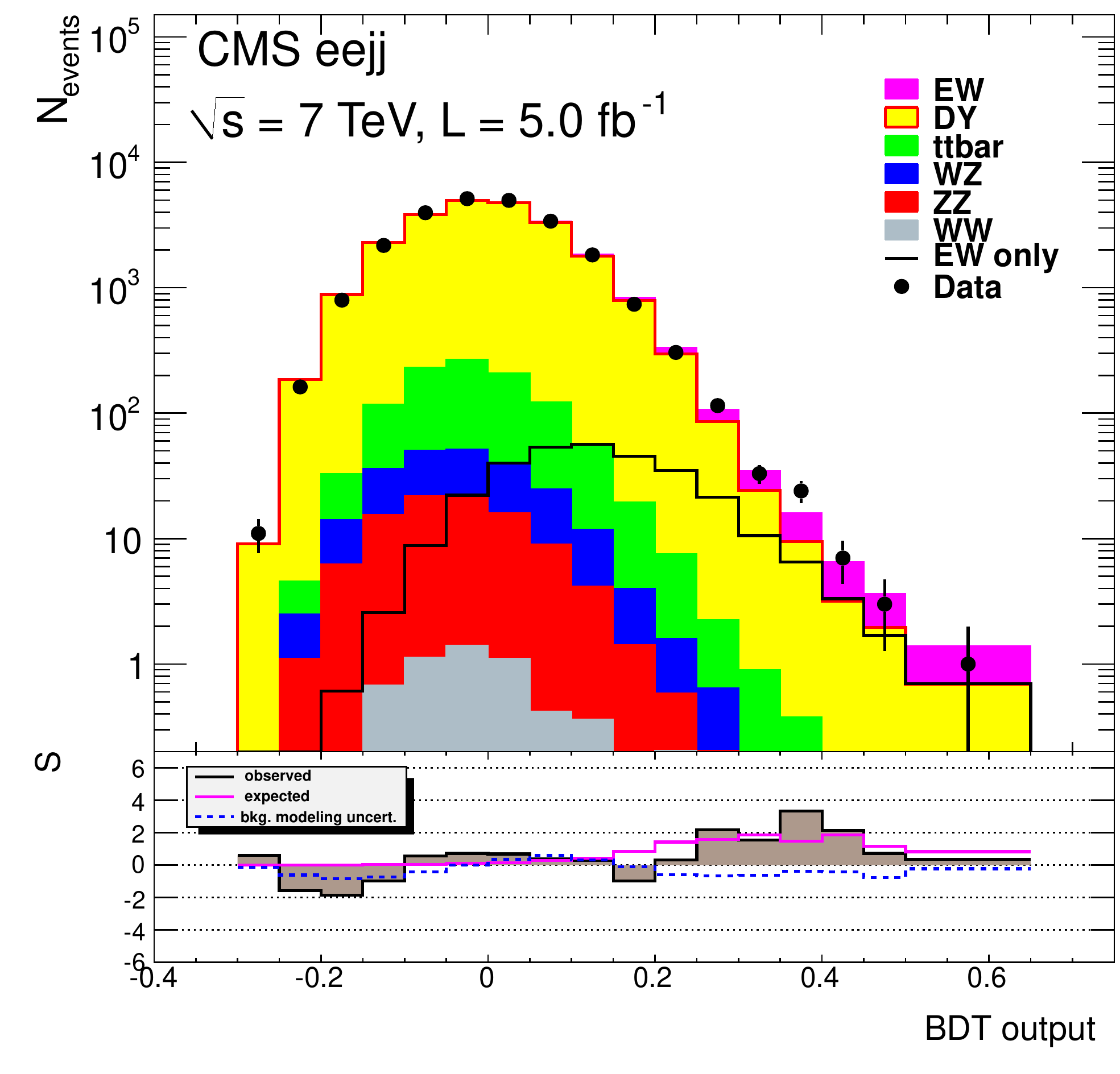}
  \includegraphics[width=0.3\columnwidth]{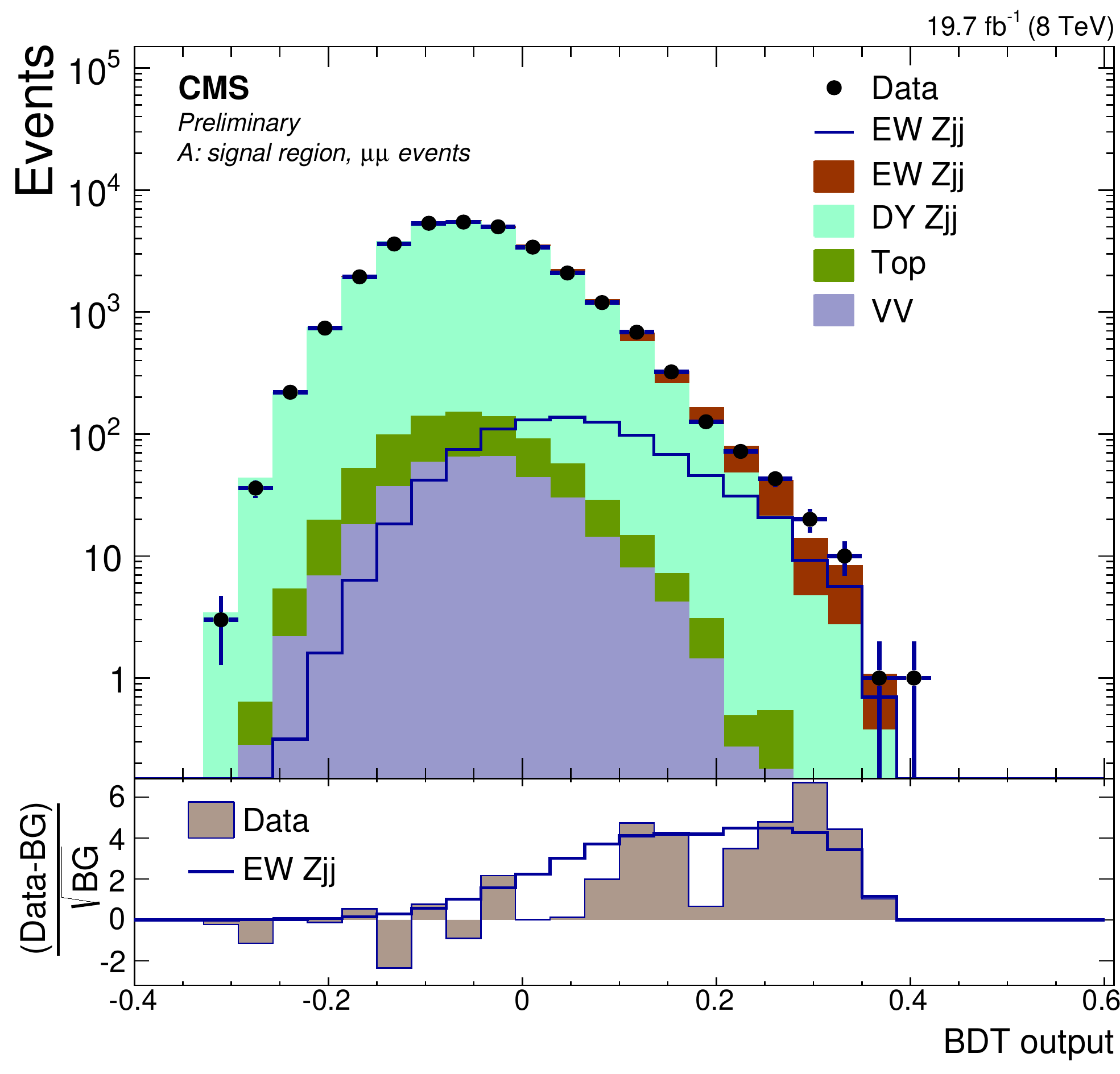}
  \includegraphics[width=0.3\columnwidth]{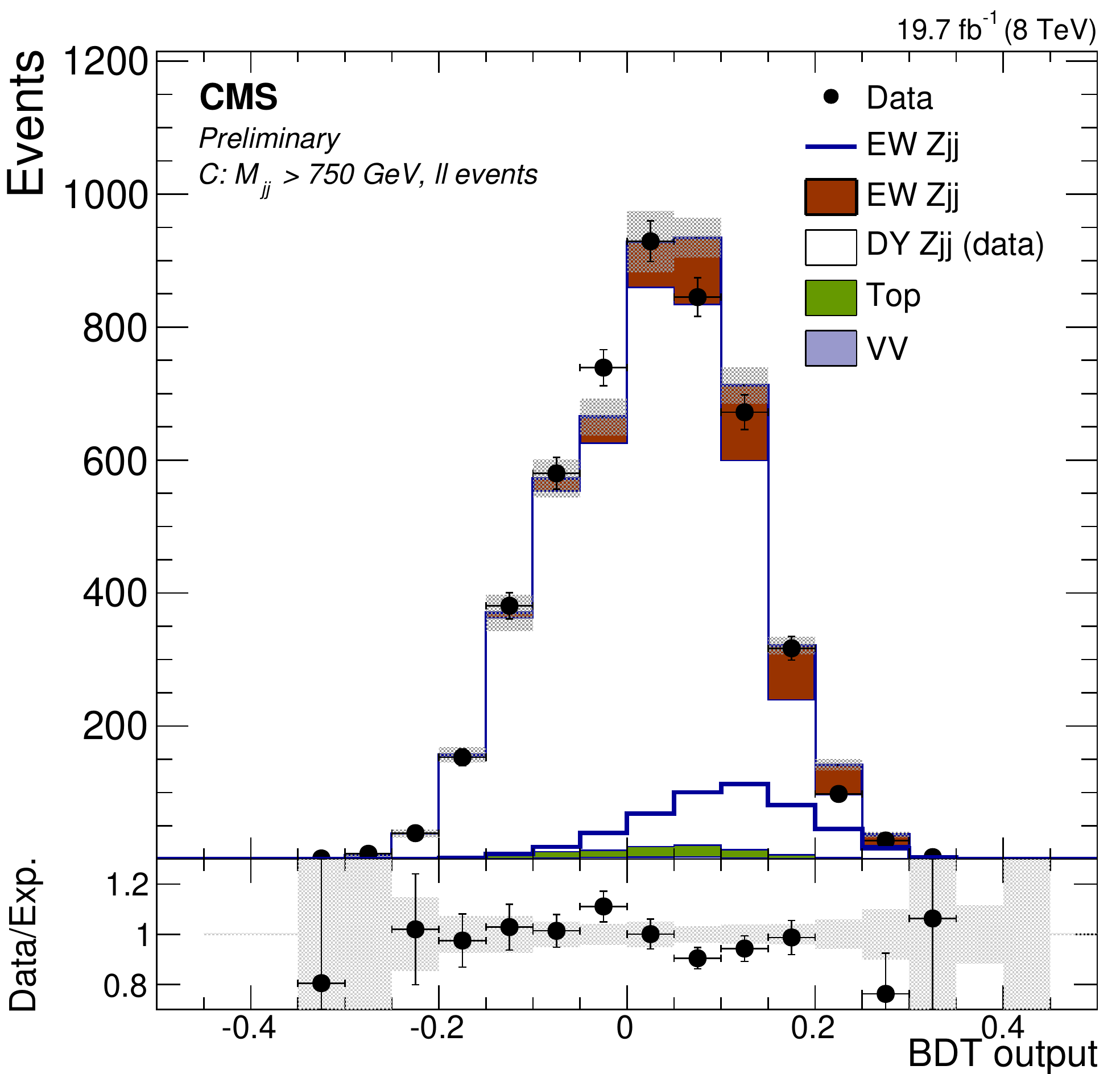}
  \caption{Shape discriminators (boosted decision trees) used for the signal cross section extraction: 7 TeV {\it(left)} and 8 TeV {\it(centre)} analysis with simulated DY background,
  8 TeV analysis with data-driven background estimation {\it(right)}.}
  \label{fig:discriminators}
\end{figure}

 To discriminate the electroweak $jjll$ signal from the backgrounds, multi variate analysis techniques are used (Figure \ref{fig:discriminators}).
 The electroweak $jjll$ cross section is extracted after fitting the data with the expected shapes for signal and background.
 After fitting for the signal strength it is extrapolated to the kinematic region $M_{ll} > 50$ GeV, $M_{jj} > 120$ GeV, $p_T^j > 25$ GeV and $|\eta_j| < 4$ (7TeV) or $|\eta_j| < 5$ (8TeV).
 The measured cross sections are $\sigma_{7TeV}^{EW\ lljj} = 154 \pm 24 \mbox{(stat)} \pm 46 \mbox{(syst)} \pm 27 \mbox{(theory)} \pm 3 \mbox{(lumi) fb}$ and 
 $\sigma_{8TeV}^{EW\ lljj} = 174 \pm 15 \mbox{(stat)} \pm 40 \mbox{(syst) fb}$, 
 and are in good agreement with the theoretical NLO predictions of 166 fb and 208 fb respectively, computed with VBFNLO.
 
 A detailed description of the input variables to the multi variate techniques, the fitting procedure and the combinations can be found in Ref. \cite{Chatrchyan:2013jya,CMS:2013qfa}.

\section{Hadronic activity in the pseudorapidity interval between the two leading jets}
\begin{figure}
  \centering
  \includegraphics[width=0.3\columnwidth]{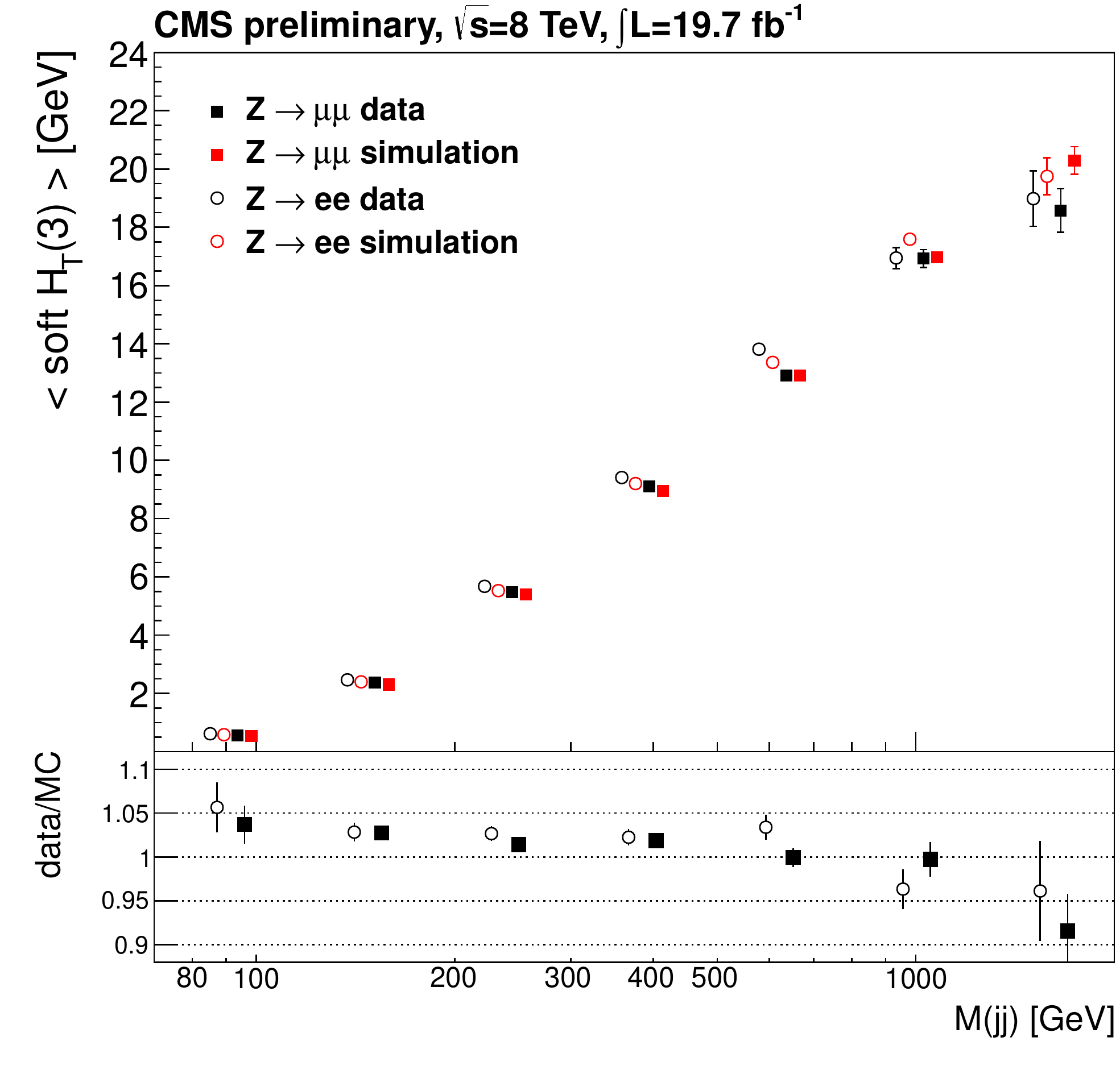}
  \includegraphics[width=0.3\columnwidth]{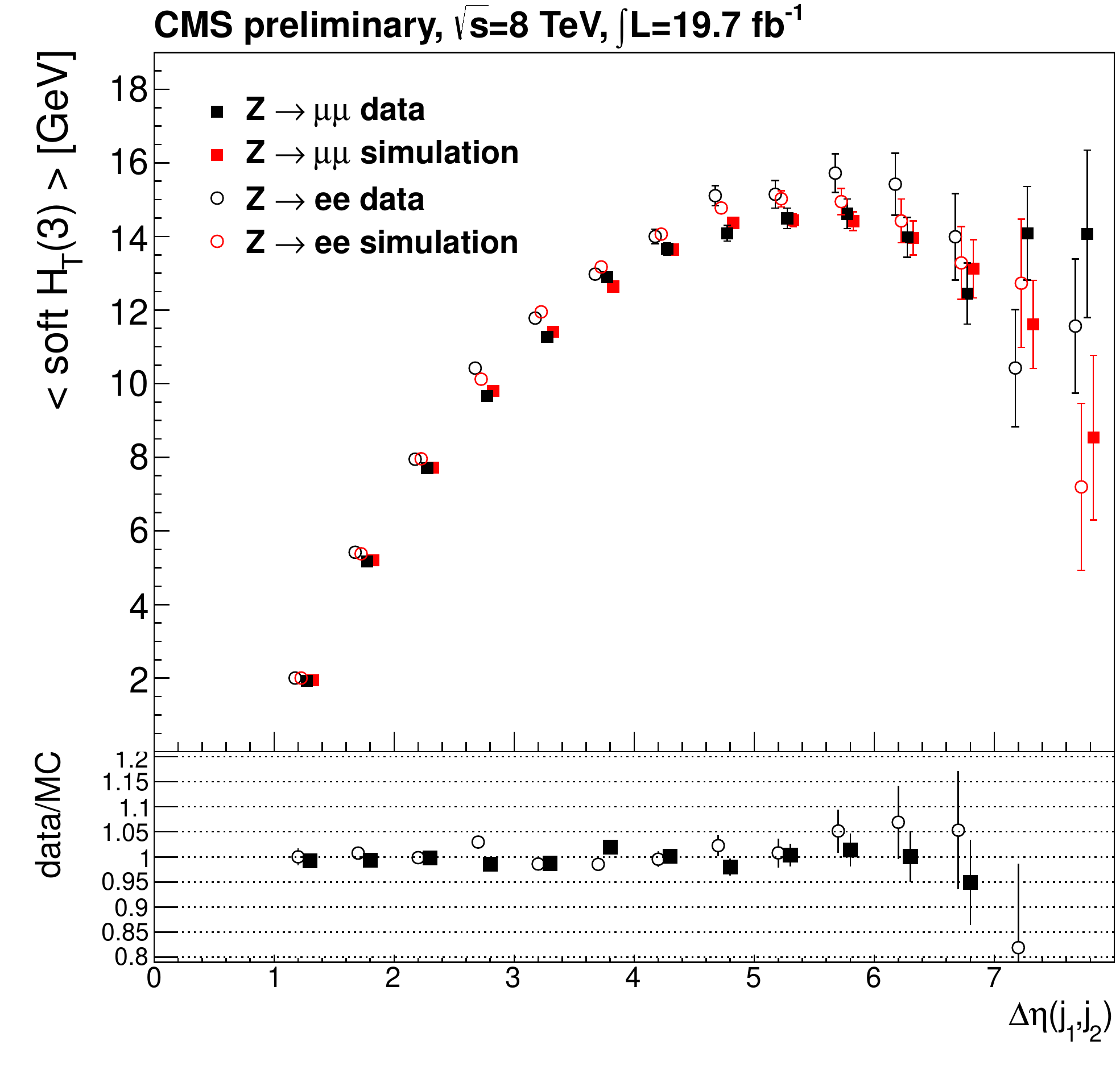}
  \caption{Average $\mathrm{H_T}$ of the three leading soft track jets in the rapidity interval between the two leading jets
  with $p_T^{j_1,j_2} > 50,30$ GeV in $Z\rm jj$ events versus {\it(left)} the dijet invariant mass {\it(right)} 
  the pseudorapidity separation between the two tagging jets.}
  \label{fig:hadronicActivity}
\end{figure}

The hadronic activity has been studied in the region between the
two leading jets ($\eta_{\text{min}}^{\text{jet}} + 0.5 < \eta < \eta_{\text{max}}^{\text{jet}} - 0.5$) using soft track-jets.
Soft track-jets are constructed using CMS ``high purity'' tracks ($\pt > 300$ MeV), associated with the main interaction primary vertex and not associated with either of the two leptons
or the two jets, which are clustered by the anti-$k_T$ algorithm with distance parameter of 0.5.
Good agreement between data and simulation is observed for the average of the three central soft track-jets versus $M_{jj}$ and $\Delta\eta_{jj}$ (Figure \ref{fig:hadronicActivity}).\\

\begin{figure}
  \centering
  \includegraphics[width=0.3\columnwidth]{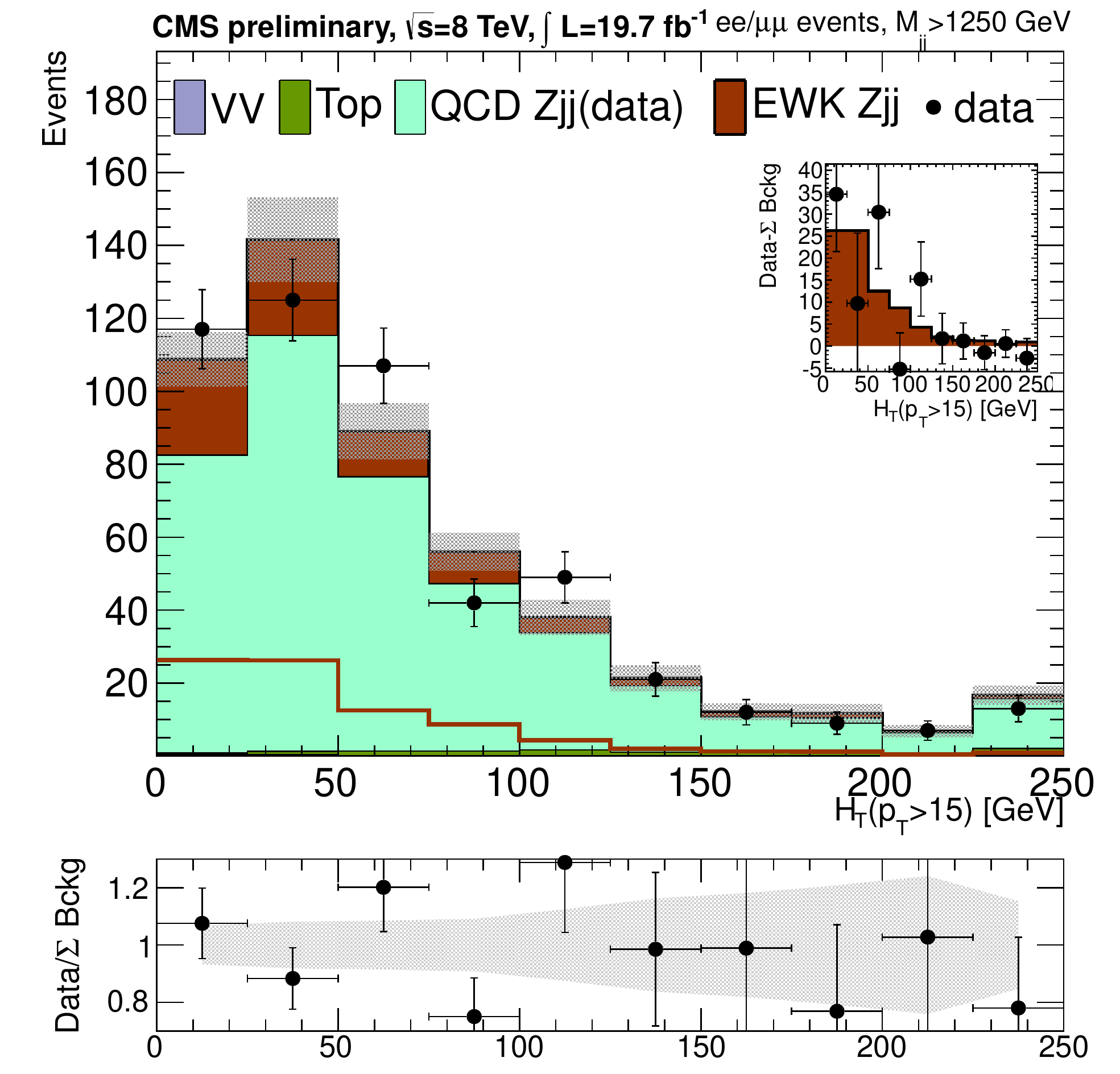}
  \includegraphics[width=0.3\columnwidth]{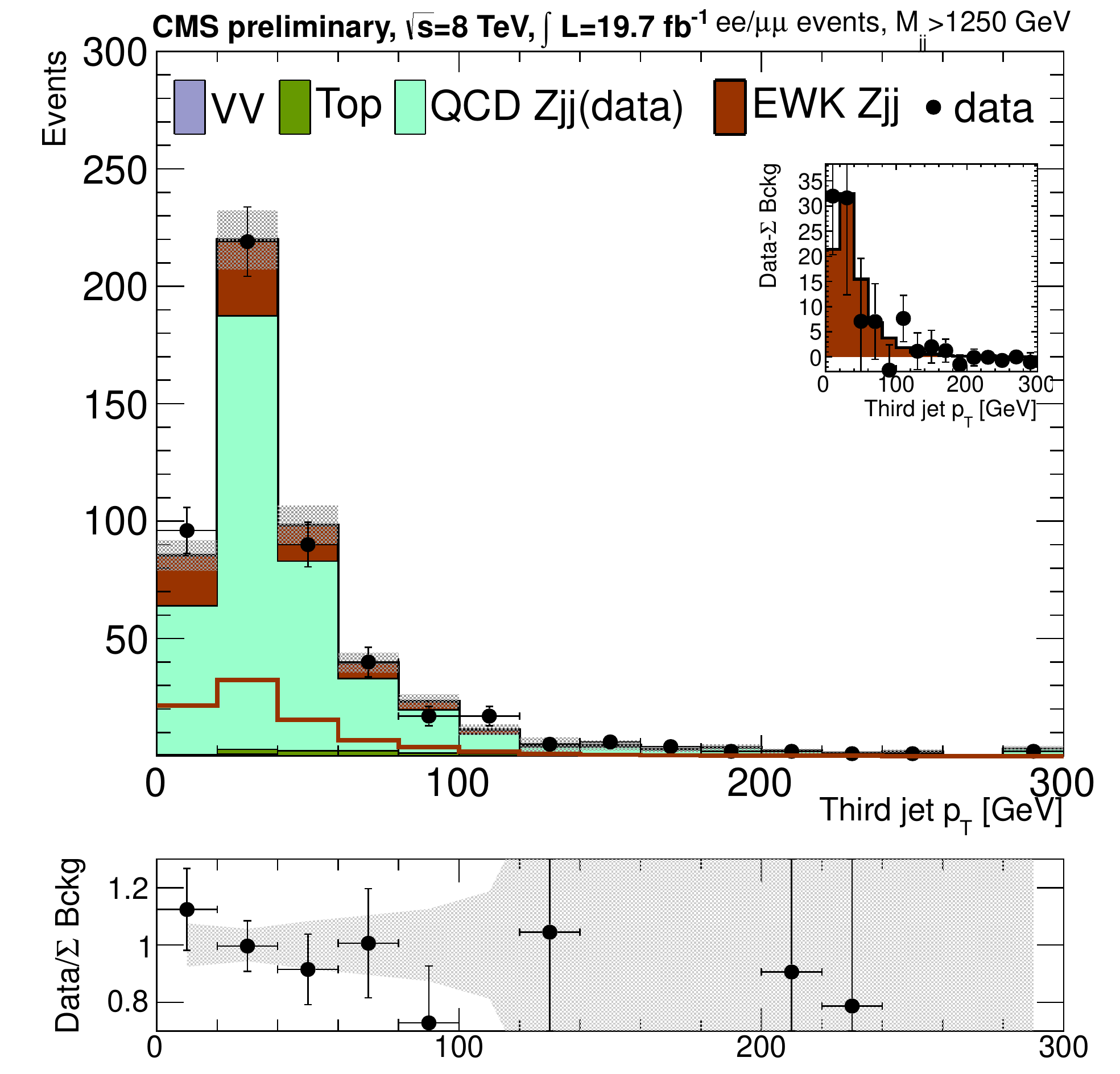}
  \includegraphics[width=0.3\columnwidth]{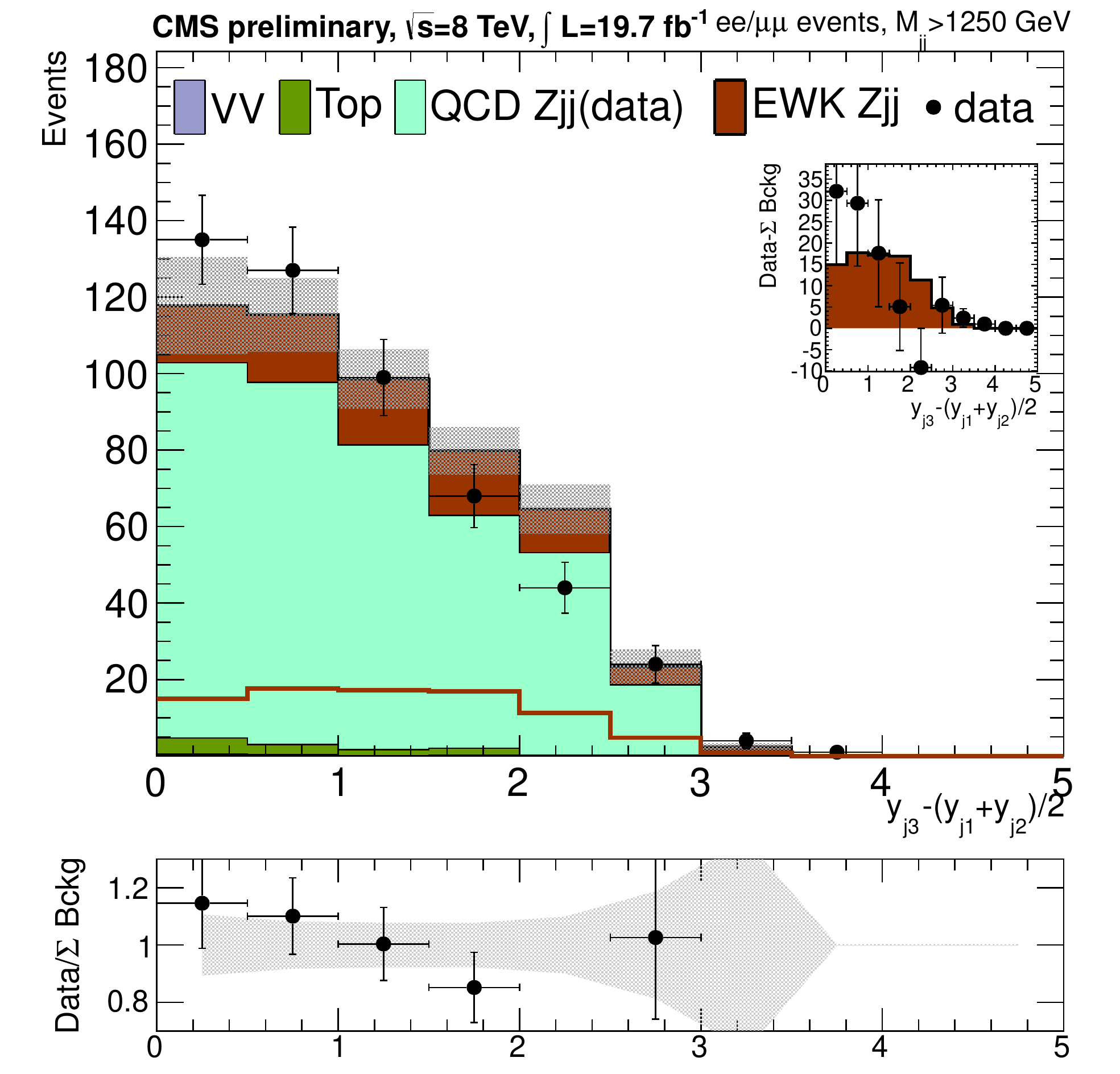}
  \caption{Control distributions for the hadronic activity for events with $M_{jj} > 1250$ GeV: scalar sum of all jets with $p_T > 15$ GeV found in the pseudorapidity distance
  between the tagging jets {\it(left)} $p_T$ of the third jet {\it(middle)} $y^*_{j_3}$  {\it(right)}.
  The background prediction is modelled from the photon control sample.}
  \label{fig:hadronicActivity2}
\end{figure}

\begin{figure}
  \centering
  \includegraphics[width=0.3\columnwidth]{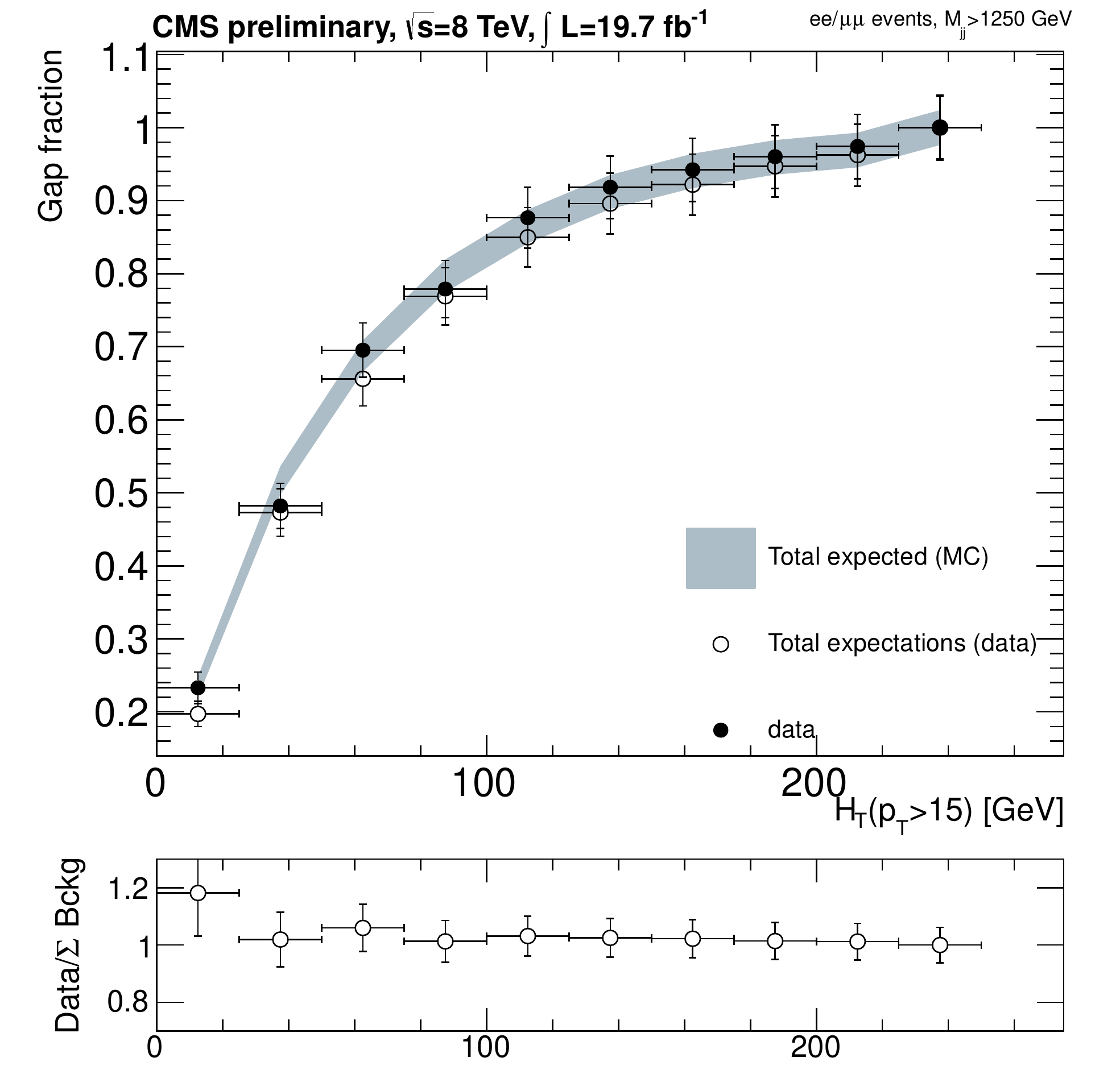}
  \includegraphics[width=0.3\columnwidth]{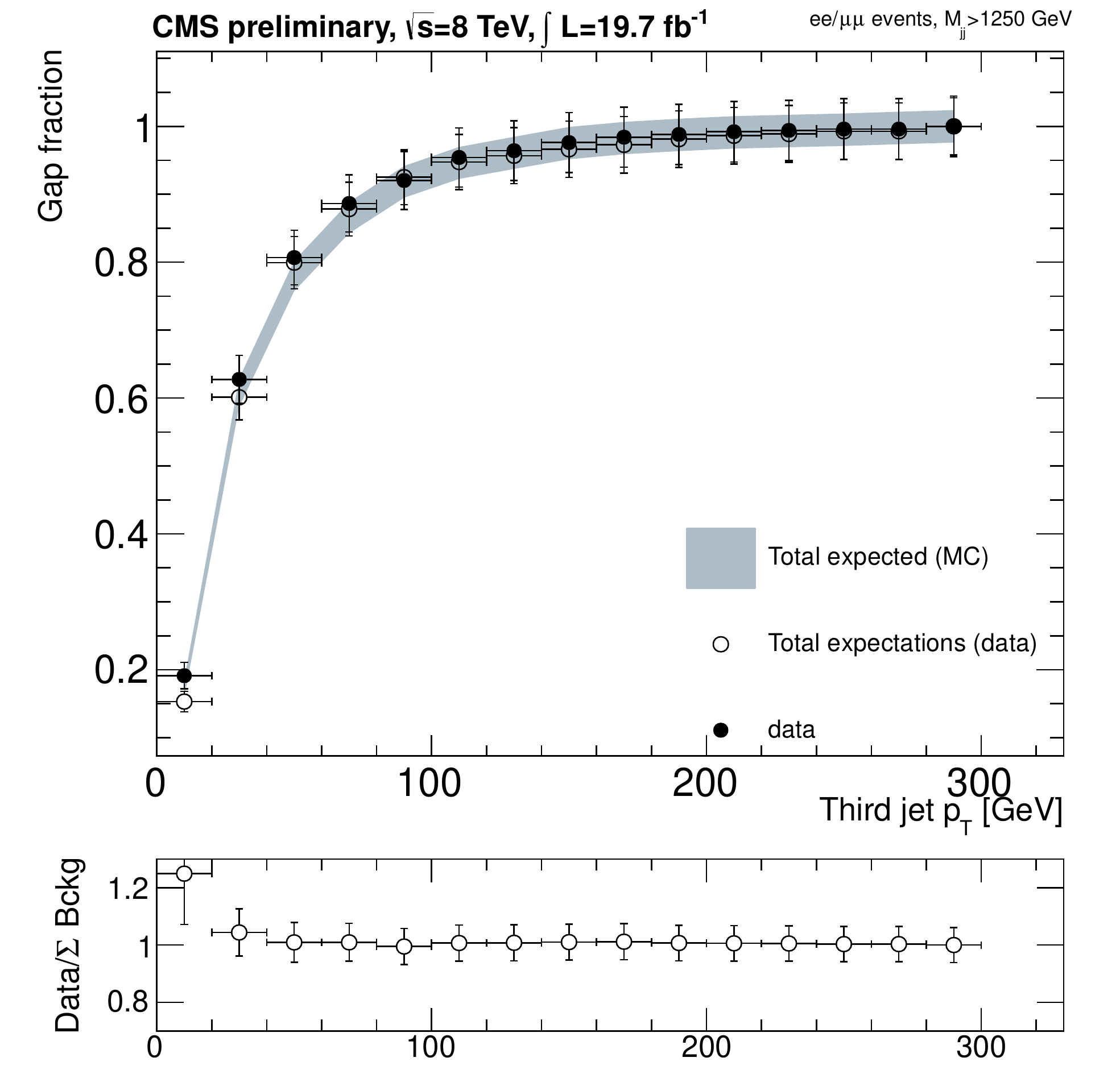}
  \caption{Gap fraction versus the scalar sum of the $p_T$ of all jets  {\it(left)} the $p_T$ of the third jet  {\it(right)}.}
  \label{fig:veto}
\end{figure}

The hadronic activity is also studied in a high purity region for which $M_{jj} > 1250$ GeV, with the use of additional jets ($\pt > 15$ GeV) found in the pseudorapidity distance between the 
two leading jets.
Figure \ref{fig:hadronicActivity2} shows the scalar sum $H_T$ of the $p_T$ of these jets, the transverse momentum of the third jet, and 
its pseudorapidity measured in the dijet rest frame. These distributions could be used to compute the efficiency of a hadronic veto either based
on the transverse momentum of the third jet or on the $H_T$ variable, as shown in Figure \ref{fig:veto}. The gap fraction corresponds to the fraction of events for which the tested variable
does not exceed a given threshold and is calculated for data, simulation and the data-driven prediction.

\section{Conclusions}
The cross section of the pure electroweak production of a Z boson in association with two jets has been measured at 7 and 8 TeV.
A data-driven background model based on a $\gamma$ plus jets sample was introduced for the 8 TeV analysis. Different studies were performed on
the hadronic activity in the rapidity interval between the two jets.


\end{document}

%% file: econfmacros.tex
\def\beq{\begin{equation}}
\def\eeq#1{\label{#1}\end{equation}}
\def\eeqn{\end{equation}}

\def\beqa{\begin{eqnarray}}
\def\eeqa#1{\label{#1}\end{eqnarray}}
\def\eeqan{\end{eqnarray}}

\let\bar=\overbar

\def\Dslash{\not{\hbox{\kern-4pt $D$}}}
\def\dslash{\not{\hbox{\kern-2pt $\del$}}}

\def\msb{{\bar{\ssstyle M \kern -1pt S}}}




%% file: LHCP_2014_-_ewkZjj.bbl
\begin{thebibliography}{99}


\bibitem{Chatrchyan:2013jya}
  CMS Collaboration,
  JHEP {\bf 1310} (2013) 062
  [arXiv:1305.7389 [hep-ex]].
  
\bibitem{CMS:2013qfa}
  CMS Collaboration,
  CMS-PAS-FSQ-12-035.



\end{thebibliography}
